\documentclass[pra,aps,twocolumn,superscriptaddress]{revtex4}
\usepackage[T1]{fontenc}

\usepackage[utf8]{inputenc}
\usepackage{amsmath,amssymb}
\usepackage[capitalize]{cleveref}
\usepackage{amsfonts}
\usepackage{epsfig,pstricks,graphicx}
\usepackage{bm}
\usepackage{color}
\def\id{{\rm 1\kern-.22em l}}
\usepackage{enumerate}

\begin{document}

\title{Experimental verification of time-order-dependent correlations in three-qubit Greenberger-Horne-Zeilinger-class states} 

\author{Artur Barasi\'nski}
\email{artur.barasinski@upol.cz}
\affiliation{RCPTM, Joint Laboratory of Optics of Palack\'{y} University and Institute of Physics of CAS, Faculty of Science, Palack\'{y} University, 17. Listopadu 12, 771 46 Olomouc, Czech Republic}

\author{Anton\'{i}n \v{C}ernoch}
\email{antonin.cernoch@upol.cz}
\affiliation{RCPTM, Joint Laboratory of Optics of Palack\'{y} University and Institute of Physics of CAS, Faculty of Science, Palack\'{y} University, 17. Listopadu 12, 771 46 Olomouc, Czech Republic}

\author{Karel Lemr}
\email{k.lemr@upol.cz}
\affiliation{RCPTM, Joint Laboratory of Optics of Palack\'{y} University and Institute of Physics of CAS, Faculty of Science, Palack\'{y} University, 17. Listopadu 12, 771 46 Olomouc, Czech Republic}

\author{Jan Soubusta}
\email{jan.soubusta@upol.cz}
\affiliation{RCPTM, Joint Laboratory of Optics of Palack\'{y} University and Institute of Physics of CAS, Faculty of Science, Palack\'{y} University, 17. Listopadu 12, 771 46 Olomouc, Czech Republic}

\begin{abstract}
In this paper, we investigate the genuine three-way nonlocality which is recognized as the strongest form of tripartite correlations. We consider theoretically and experimentally a series of suitable Bell-type inequalities a violation of which is sufficient for the detection of three-way nonlocality. For the generalized GHZ (gGHZ) states, it is demonstrated that they do violate tripartite Bell-type inequalities for any degree of tripartite entanglement even if they do not violate Svetlichny inequality. It implies that three-way entangled gGHZ can always exhibit genuine three-way nonlocality under the requirement of time-order-dependent principle.
Furthermore, we have determined the maximal amount of noise admissible for the gGHZ states to still remain genuine three-way nonlocal.
\end{abstract}

\maketitle

\section{Introduction}

A fundamental problem in physics is to identify which correlations among different events can be observed within the description based on quantum mechanics. It is known that quantum theory allows correlations between spatially separated systems that are fundamentally different from classical counterparts. Moreover, even a very general quantum theory imposes nontrivial constraints on the allowed correlations among remote observers and hence, the answer for this question becomes not trivial.

One of the most important and at the same time intriguing classes of quantum correlations is known as nonlocal correlations, which has been recognized as an essential resource for quantum information tasks \cite{Brunnerrmp86_2014}, such as quantum key distribution \cite{Bennet_proc, Ekertprl67_1991}, communication complexity \cite{Buhrmanrmp82_2010}, and randomness generation \cite{Pironionature464_2010}. Specifically, the correlations between outcomes of measurements on two or more spatially separated subsystems are said to be nonlocal, if they cannot be reproduced by any local-hidden variable (LHV) model \cite{Bellphysics1_1964}. In particular, for the tripartite case, disused in this paper, the standard definition of nonlocality according to Bell \cite{Bellphysics1_1964} can be written as
\begin{equation}
P(abc|XYZ) =  \sum_{\lambda} q(\lambda) P_{\lambda}(a|X) P_{\lambda}(b|Y) P_{\lambda}(c|Z),
\end{equation}
where local maps $P_{\lambda}(\delta|\Delta)$ are one-partite probability distribution of obtaining outcome $\delta=\{a,b,c\}$ with respect only to the local measurement $\Delta=\{X,Y,Z\}$ and the past variables $\lambda$. Naturally, the hidden variable $\lambda$ is distributed according to $q(\lambda)$ fulfilling requirements $0 \leq q(\lambda) \leq 1$ and $\sum_{\lambda} q(\lambda) = 1$. If such factorization of joined probability distribution  exists then the correlations described by $P(abc|XYZ)$ can be explained by locally causal theory. Otherwise, they are referred to as nonlocal correlations. 

This definition of nonlocality and resulting Bell-type inequalities, although sufficient for bipartite systems \cite{Bartkiewiczpra88_2013,Bartkiewiczpra97_2018,Bartkiewiczpra95_2017}, do not cover all possible variants of multipartite nonlocal correlations. As in the case of quantum entanglement, nonlocal correlations display a much richer and more complex structure for the multipartite case than the bipartite one. For instance, when restricting our consideration to the case of three observers, it is possible to have a hybrid local-nonlocal scenario in which (any) two parties may share bipartite nonlocal probability distribution. According to Svetlichny \cite{Svetlichnyprd35_1987}, an appropriate hybrid local-nonlocal model is given by
\begin{equation}
P(abc|XYZ) = \sum^3_{i=1}  \sum_{\lambda_i} q(\lambda_i) P_{\lambda_i}(\delta_i|\Delta_i) P_{\lambda_i}(\delta_j \delta_k|\Delta_j \Delta_k),
\label{eq:hybrid_model}
\end{equation}
where $\{i,j,k\}$ is an even permutation of $\{1, 2, 3\}$, $\delta = \{a,b,c\}$, and $\Delta =\{X,Y,Z\}$. As previously, we expect that $0 \leq q(\lambda_i) \leq 1$ and $\sum^3_{i=1} \sum_{\lambda_i} q(\lambda_i)=1$. 

It turns out that quantum-mechanical description of nature authorizes the existence of correlations that cannot be explained by these more general LHV models and hence, those correlations exhibit genuine multipartite or n-way nonlocality \cite{Collinsprl88_2002b,Bancalprl106_2011}. To test for such type of nonlocality, the Svetlichny inequality (a certain type of Bell-type inequality) is usually used \cite{Collinsprl88_2002b, Bancalpra88_2013}
\begin{eqnarray}
I_{185} = \frac{1}{4} [&-&\langle A_0B_0C_0\rangle - \langle A_1B_0C_0\rangle + \langle A_0B_1C_0\rangle
\nonumber\\
&-& \langle A_1B_1C_0\rangle - \langle A_0B_0C_1\rangle + \langle A_1B_0C_1\rangle \nonumber\\
&-& \langle A_0B_1C_1\rangle- \langle A_1B_1C_1\rangle] \leq 1,
\label{eq:I_185}
\end{eqnarray}
where $A$, $B$, and $C$ denote the three particles, $\langle A_iB_jC_k\rangle$ means correlation coefficient and subscripts $0$, $1$ denote two different measurements. Note that the inequality $I_{185}$ presented here has been normalized which will be explained later.

The above-described phenomenon has naturally attracted much interest since the genuine multipartite nonlocality represents the strongest form of multipartite nonlocality, in which nonlocal correlations are established among all the parties of the system \cite{Lavoienjp11_2009,Lupra84_2011a,Lupra84_2011b,Hamelnatutephot8_2014,Adessoprl112_2014, Chenprl112_2014}. In particular, it has been shown \cite{Emarypra69_2004,Ghoseprl102_2009} that within the three-parameter pure GHZ subfamily, $|\Phi(\theta, \xi, \omega) \rangle = \cos \theta~|000\rangle_{ABC} + \sin \theta~|1\rangle_A \otimes (\cos \xi |0\rangle_B + \sin \xi |1\rangle_B) \otimes (\cos \omega |0\rangle_C + \sin \omega |1\rangle_C)$, the upper and lower bounds of maximal violation of Svetlichny inequality for a given tripartite entanglement monotony (the three-tangle \cite{CKWpra61_2000}) are provided by the maximal slice (MS) states \cite{Carteretjpa33_2000} and the generalized GHZ (gGHZ) states \cite{Durpra62_2000}, respectively. Both states are given by $|\Phi(\pi/4, \pi/2, \omega)\rangle$ and $|\Phi(\theta, \pi/2, \pi/2)\rangle$, respectively. For a general case of pure three-qubit state, the upper bound is given by the tetrahedral states which are recognized as a lower bound of the primary yield of GHZ states from the infinitesimal distillation protocol \cite{Barasinskisr8_12305_2018}. 

The hierarchy which occurs between these three states causes an important consequence in the experimental detection of genuine multipartite nonlocal correlations where the inherent presence of noise must be taken into account, i.e., the more the Bell-type inequality is violated the greater amount of noise is needed to suppress nonlocality. In other words, the more the Bell-type inequality is violated, the more the nonlocality of the state is robust against noise. On the other hand, it is also very interesting that not every pure tripartite entangled state reveals Svetlichny correlations \cite{Barasinskisr8_12305_2018}. For example, the gGHZ state admits the hybrid local-nonlocal model in Eq. \eqref{eq:hybrid_model} when $\theta \leq 22.5^{\circ}$. 

The explanation of the last observation is based on the fact that the hybrid local-nonlocal model proposed by Svetlichny allows for correlations capable of two-way signaling among some parties. As a result, Svetlichny’s notion of genuine nonlocality is inconsistent with the operational approaches \cite{Gallegoprl109_2012}. To overcome this problem an alternative definition of genuine multipartite nonlocality based on time-order-dependent correlation (in literature denoted as TOBL) has been defined \cite{Gallegoprl109_2012,Bancalpra88_2013}. Specifically, it is assumed that $P_{\lambda_i}(\delta_j \delta_k|\Delta_j \Delta_k)$ in Eq. \eqref{eq:hybrid_model} is at most one-way signaling.
This model of nonlocality is crucial for the simulation of quantum correlations in all protocols where measurements performed on a particular system may depend on the measurement outcome obtained from another system, e.g. measurement-based computation. In such cases, Svetlichny-type simulation models in which all measurements settings are given at the same time are not relevant.

The hybrid local-nonlocal model can be further extended to the nonsignaling principle \cite{Barrettpra71_2005, Almeidapra81_2010, Bancalpra88_2013}, in which the correlations are nonsignaling for all observers. 
Consequently, the existence of three-way nonlocality is certified by the violation of at least one of 185 inequalities \cite{Bancalpra88_2013}, where the Svetlichny's original inequality is only one of them.

For that reason an important question arises whether previous observations are still valid if one considers other types of LHV models. In particular, (i) what kind of correlations (i.e., Svetlichny correlations, TOBL, nonsignalling correlations) are needed to reveal three-way nonlocal correlations for all gGHZ states and hence, to satisfy the Gisin’s theorem \cite{Gisinpla154_1991, Gisinpla162_1992}; (ii) Is the relation between gGHZ state and MS states still valid if one assumes more general LHV model?; (iii) What is the maximum value of noise required for the non-maximally entangled gGHZ states to produce each type of three-way nonlocality.

Note that partial answers for the first and the last questions are given in Refs. \cite{Mukherjeejpa48_2015} and \cite{Bancalpra88_2013}, respectively. Here, we present an extensive theoretical and experimental analysis in that field, taking into consideration all 185 Bell-type inequalities. To confirm our predictions, we use experimental setup based on correlated photon pairs generation in cascade of two BBO crystal via the process of spontaneous parametric down-conversion \cite{Kwiat_PRA60, White_PRL83, Halenkova_AO51}. 
In our experiment, two qubits are encoded into polarization and one qubit is encoded into propagation path. We demonstrate the ability of fast and efficient detection of three-way nonlocality, a task that was not possible with previous three-photon sources \cite{Lavoienjp11_2009,Lupra84_2011a,Lupra84_2011b}. Therefore, our experiment opens the  door for future investigations of genuine multipartite nonlocal correlations based on the recent concept of nonlocal fraction \cite{Fonsecapra92_2015, Rosierpra96_2017, Lipinskanpj_2018, Barasinskipra98_2018, Fonsecapra98_2018}, where accumulation of a large amount of data is required.

\section{Nonlocality of GHZ states under the operational principle}
\label{Sec2}

\begin{figure}
\centering
\includegraphics[width=8.5cm]{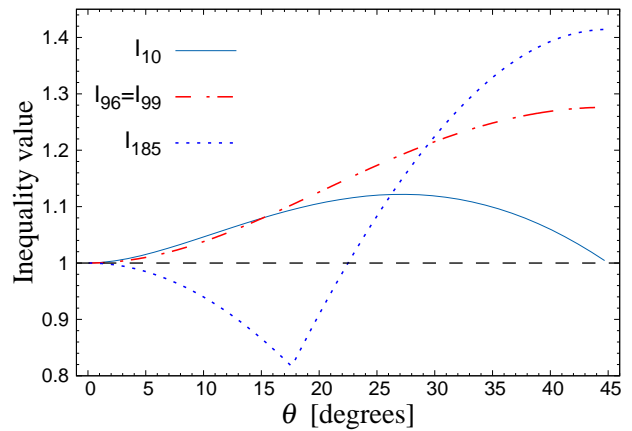}
\caption{Theoretical dependence of normalized Bell-type inequalities ($I^{\max}_{10}, I^{\max}_{96}, I^{\max}_{99},I^{\max}_{185}$) for the gGHZ state as a function of angle $\theta$. Dashed line represents the classical limit.
}
\label{fig:pure_states_CT}
\end{figure}

In this paper we discuss the following problem: Suppose we have three observers Alice (A), Bob (B), and Charlie (C) who perform dichotomic measurements on a quantum state $\rho$ and each party has a choice of two measurement settings. Moreover, we assume that $\rho = p |\psi\rangle \langle \psi| + \frac{1-p}{8} \id$, where $|\psi\rangle$ denotes a tripartite pure state, $\id$ stands for identity matrix and $0 \leq p \leq 1$. Suppose also that we have a given Bell-type operator $\hat{I}$ (e.g., Svetlichny operator \cite{Ghoseprl102_2009}). Then, 
the nonlocal correlations of such a system can be generated if the expectation value for $\rho$ exceeds the Bell-type inequality
\begin{eqnarray}
I(\rho) = \textrm{Tr}(\hat{I}\rho) = p \textrm{Tr}(\hat{I}|\psi\rangle \langle \psi|)\leq C_{LHV},
\label{eq:Imax_rho}
\end{eqnarray}
where the second equality holds because the operator $\hat{I}$ is traceless. The parameter $C_{LHV}$ depicts the upper threshold of one inequality for local realism \cite{Bancalpra88_2013}. To render the discussion more transparent, in the rest of the paper we use Bell-type operator $\hat{I}$ normalized in such a way that $C_{LHV} \equiv 1$. 
On the basis of that result we can determine the minimal value of the $p$ parameter (maximal strength of noise) required for the quantum state to exhibit genuine multipartite nonlocality, namely
\begin{eqnarray}
p > p^{\min}(|\psi\rangle)=\frac{1}{I^{\max}(|\psi\rangle)},
\end{eqnarray}
where $I^{\max}(|\psi\rangle) = \max_{\mathcal{M}} \textrm{Tr}(\hat{I}|\psi\rangle \langle \psi|)$ is the maximal violation of the Bell-type inequality with respect to $|\psi\rangle$ and the maximum is taken over all possible measurement settings. Naturally, for various $I$,  different values of $p^{\min}$ can be reached. Therefore, in order to investigate the maximal extent to which the gGHZ and MS states can produce each type of multipartite nonlocality, we consider all $185$ Bell-type inequalities which describe the tripartite polytope of three-way correlations formulated in Ref. \cite{Bancalpra88_2013}.

As an example of Bell-type inequality, let us now consider the 96th facet inequality, a violation of which implies that nonlocal correlations are time-order dependent \cite{Bancalpra88_2013},
\begin{eqnarray}
I_{96} &=& \frac{1}{6} [2 \langle A_0B_0\rangle - \langle C_0\rangle - \langle A_0C_0\rangle - \langle B_0C_0\rangle \nonumber\\
&+& \langle A_0B_0C_0\rangle - 2 \langle A_1B_1C_0\rangle - \langle C_1\rangle + \langle A_0C_1\rangle \nonumber\\
&+& \langle B_0C_1\rangle+ \langle A_0B_0C_1\rangle - 2 \langle A_1B_1C_1\rangle] \leq 1,
\label{eq:ineq96}
\end{eqnarray}
where we assume to have an ensemble of projective measurement $A_0=\vec{a}_0 \cdot \vec{\sigma}_A$, or $A_1=\vec{a}_1 \cdot \vec{\sigma}_A$ on qubit $A$, $B_0=\vec{b}_0 \cdot \vec{\sigma}_B$, or $B_1=\vec{b}_1 \cdot \vec{\sigma}_B$ on qubit $B$, $C_0=\vec{c}_0 \cdot \vec{\sigma}_C$, or $C_1=\vec{c}_1 \cdot \vec{\sigma}_C$ on qubit $C$. All vectors $\vec{a}_0$, $\vec{a}_1$, $\vec{b}_0$, $\vec{b}_1$ and $\vec{c}_0$, $\vec{c}_1$ are of length $1$ and $\vec{\sigma}=\{\sigma_x,\sigma_y,\sigma_z\}$, where $\sigma_{x}$, $\sigma_{y}$, and $\sigma_{z}$ denote the Pauli operators associated with three orthogonal directions. 
For the generalized GHZ state,
\begin{equation}
|\mathcal{G}\rangle = \cos \theta |000\rangle_{ABC} + \sin \theta |111\rangle_{ABC},
\label{eq:ghz}
\end{equation}
where $0\leq \theta \leq \frac{\pi}{4}$, the maximum value of $I_{96}$ in Eq. \eqref{eq:ineq96} becomes
\begin{equation}
I^{\max}_{96}(|\mathcal{G}\rangle) = \frac{1}{3} [1 + 2 \sqrt{1+ \sin^2(2 \theta)}].
\label{eq:I96_solution}
\end{equation}
To achieve $I^{\max}_{96}$, one can measure with the following set of unit vector: 
$\vec{a}_0$ and $\vec{b}_0$ are aligned along the $-\vec{z}$ direction, $\vec{a}_1$ and $\vec{b}_1$ are aligned along $\vec{x}$, and vectors $\vec{c}_0$ and $\vec{c}_1$ lie in the $x-z$ plane. Specifically, $\vec{c}_i = \{-\sin \alpha,0,(-1)^i \cos \alpha \}$, where $\alpha = \arctan[\sin(2 \theta)]$ (see Appendix \ref{appendixA}).

\begin{figure}
\centering
\includegraphics[width=\columnwidth]{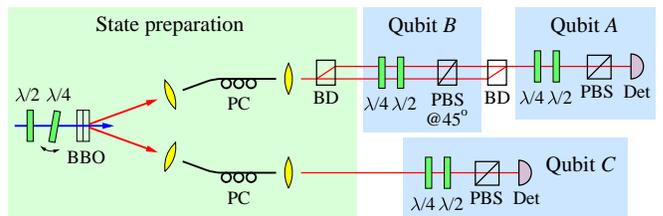}
\caption{Scheme of the experimental setup for gGHZ state preparation and three-qubit projection. 
  $\lambda/2$: half-wave plate; $\lambda/4$: quarter-wave plate; BBO: double crystal cascade; PC: polarization controller; BD: beam displacer; PBS: polarizing beam splitter; Det: detector.
  \label{fig:setup}}
\end{figure}

Interestingly, the same boundary can be found for the 99th facet inequality
\begin{eqnarray}
I_{99} &=& \frac{1}{3} [\langle A_1B_1\rangle + \langle B_1C_0\rangle + \langle A_1C_1\rangle \nonumber\\&+& \langle A_0B_0C_0\rangle - \langle A_0B_0C_1\rangle] \leq 1,
\label{eq:ineq99}
\end{eqnarray}
where all symbols have the same meaning as before. Indeed, when 
$\vec{a}_0$ and $\vec{b}_0$ are aligned along the $\vec{x}$ direction, and $\vec{a}_1$ and $\vec{b}_1$ are aligned along $\vec{z}$ and $\vec{c}_i= \{(-1)^i \sin \alpha,0,\cos \alpha \}$ then, one has $I^{\max}_{96}(|\mathcal{G}\rangle) \equiv I^{\max}_{99}(|\mathcal{G}\rangle)$ (see also Ref. \cite{Mukherjeejpa48_2015}).

As we see in Fig. \ref{fig:pure_states_CT}, when $\theta < 29.5^{\circ}$ the normalized 96th facet inequality can be violated stronger than the normalized Svetlichny inequality $I^{\max}_{185}(|\mathcal{G}\rangle) =\max \big\{\sqrt{1-\sin^2(2 \theta)},\sqrt{2 \sin^2(2 \theta)} \big \}$ (discussed in details in Ref. \cite{Ghoseprl102_2009}). This means that for nonmaximally entangled gGHZ states, there exists a range of $p$ values for which these states are  too noisy to exhibit Svetlichny nonlocality, but can still generate the time-order-dependent correlations.
This result can be enhanced even more if one takes the 10th facet inequality (also TOBL class inequality) into consideration:
\begin{eqnarray}
I_{10} &=& \frac{1}{6} [-2 \langle A_1\rangle - \langle B_0\rangle + \langle A_1B_0\rangle - \langle B_1\rangle + \langle A_1B_1\rangle \nonumber\\
&-& \langle C_0\rangle + \langle A_1C_0\rangle + \langle B_0C_0\rangle - \langle A_0B_0C_0\rangle \nonumber\\
&+& 2 \langle A_1B_0C_0\rangle + \langle A_0B_1C_0\rangle + \langle A_1B_1C_0\rangle
\nonumber\\
&-& \langle C_1\rangle + \langle A_1C_1\rangle + \langle A_0B_0C_1\rangle + \langle A_1B_0C_1\rangle \nonumber\\
&+& \langle B_1C_1\rangle - \langle A_0B_1C_1\rangle + 2 \langle A_1B_1C_1\rangle].
\end{eqnarray}    
Then, the maximal violation of $I_{10}$ with respect to the gGHZ state can be approximated as
\begin{equation}
I^{\max}_{10}(|\mathcal{G}\rangle) \approx 1 + 0.0622~\theta + 1.697~\theta^2 - 3.391~\theta^3 + 1.442~\theta^4.
\label{eq:I10_solution}
\end{equation}
Note that the analytical solution, although attainable for a given $\theta$, is too complicated to be presented here (see Appendix \ref{appendixB}). Direct comparison of $I^{\max}_{10}(|\mathcal{G}\rangle)$ and $I^{\max}_{96}(|\mathcal{G}\rangle)$ implies that the first one corresponds to the slightly higher threshold visibility $p^{\min}$ than the former inequality when $\theta<14.94^{\circ}$ (see Fig. \ref{fig:pure_states_CT}). Furthermore, numerical optimization performed on the remaining Bell-type inequalities given in Ref. \cite{Bancalpra88_2013} reveals that the above-described results are sufficient to estimate the threshold visibility of the gGHZ states. Consequently, the higher threshold visibility $p^{\min}$ for the gGHZ states can be summarized as
\begin{equation}
p^{\min}(|\mathcal{G}\rangle)= \left\{ \begin{array}{ll}
\frac{1}{I^{\max}_{10}(|\mathcal{G}\rangle)} & \textrm{for $0\leq \theta < 14.94^{\circ}$}\\
\frac{1}{I^{\max}_{96}(|\mathcal{G}\rangle)} & \textrm{for $14.94^{\circ}\leq \theta < 29.5^{\circ}$}\\
\frac{1}{I^{\max}_{185}(|\mathcal{G}\rangle)} & \textrm{for $29.5^{\circ}\leq \theta < 45^{\circ}$}.
\end{array} \right.
\label{eq:pminGHZ}
\end{equation}

\begin{figure}
\centering
\includegraphics[width=.9\columnwidth]{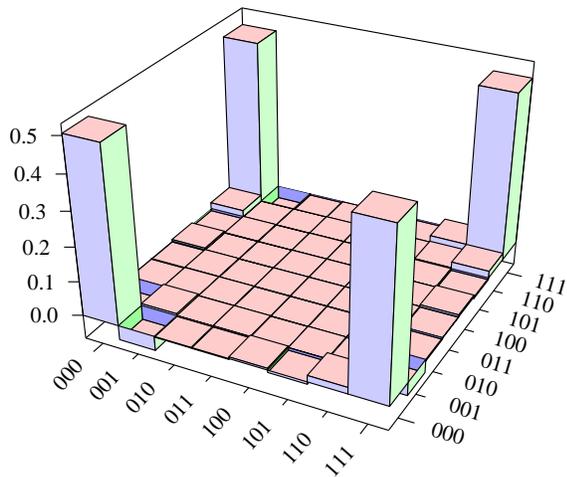}
\caption{Real part of the density matrix $\rho_{\rm exp}$ of the generated gGHZ state with $\theta =  45^{\circ}$. Imaginary parts of the matrix components are negligible.
  \label{fig:tomography_GHZ}
}
\end{figure}

At the end of this section, let us mention that similar calculations have been made for the MS state. In that  case we have found that Svetlichny correlations are sufficient to determine minimal values of $p$, i.e., $I^{\max}_i (|MS\rangle) < I^{\max}_{185}(|MS\rangle) = \sqrt{1+\sin^2 \omega}$ for $i=1,2,\dots,184$. 
In order to compare the threshold visibility achieved for gGHZ and MS state, we use a quantity called three-tangle $\tau$ \cite{CKWpra61_2000} which refers to tripartite entanglement, where $\tau(|\mathcal{G}\rangle)=\sin^2(2 \theta)$ and $\tau(|MS\rangle)=\sin^2 \omega$. Then, one can find $p(|MS\rangle) > p(|\mathcal{G}\rangle)$ if and only if $\tau > 0.0434$. In other words, the three-way nonlocal correlations generated with gGHZ states are more fragile against noise then correlation produced with MS states almost in the entire range of tripartite entanglement.

\section{Experimental Setup}

We use a standard configuration for generation of photon pairs in the process of spontaneous parametric down-conversion; see Fig.~\ref{fig:setup}. A cascade of two BBO crystals is used in a so-called Kwiat configuration to generate polarization-correlated photon pairs \cite{Kwiat_PRA60, White_PRL83, Halenkova_AO51}. The beam displacer (BD) in the state preparation part allows one to split horizontal ($H$) and vertical ($V$) components of the first photon state to two spatial modes ($0, 1$):
\begin{equation}
  \cos\theta |0HH \rangle + \sin\theta |1VV \rangle,
\end{equation}
where $\{0,1\}$ labels the first photon's spatial mode and $\{H, V\}$ the polarization state of the first and second photon, respectively. Associating polarization states $\{H, V\}$ with logical state $\{0,1\}$ immediately identifies the state as $|\mathcal{G}\rangle$ according to Eq.~(\ref{eq:ghz}).
The angle $\theta$ and zero phase between the two components of the state (13) are set using rotation and tilt of two wave plates in the pump laser beam.
To measure any correlation coefficient $\langle A_iB_jC_k\rangle$ we set accordingly the six wave plates in the projection part of the setup and measure coincidence counts on the two detectors, denoted $f^{+++}$. For normalization we have to measure also all seven orthogonal projections, i.e., $f^{-++}$, which means orthogonal projection in qubit $A$ and likewise for other terms. Correlation function is then calculated as
\begin{equation}
  \langle A_iB_jC_k\rangle = 
  {S_{ABC}^+ - S_{ABC}^- \over S_{ABC}^+ + S_{ABC}^-},
  \label{eq:corrFun}
\end{equation}
where
\begin{eqnarray*}
  S_{ABC}^+ &=& f^{+++} + f^{+--} + f^{-+-} + f^{--+}, \\
  S_{ABC}^- &=& f^{---} + f^{-++} + f^{+-+} + f^{++-}.
\end{eqnarray*}

\begin{table}
\caption{Characteristic of experimentally generated states. First column refers to angle $\theta$ defined in Eq. \eqref{eq:thetaM} while $\theta_{opt}$ (second column) corresponds to angle which maximize fidelity $F$ given in Eq. \eqref{eq:fidelity_opt}. $P$ depicts purity, $F_{\max}$ stands for an upper bound on fidelity. $N^{\textrm{exp}}_{\textrm{tri}}$ and $N_{\textrm{tri}}(\theta,P)$ denote tripartite negativity for experimentally generated states and the theoretical model discused in Appendix \ref{appendixC}, respectively.
}
\label{tab:tab1}
\begin{ruledtabular}
\begin{tabular}{c c c c c c c }
 $\theta$ & $\theta_{opt}$ & $F$ & $P$ & $F_{\max}$ & $N^{\textrm{exp}}_{\textrm{tri}}$ & $N_{\textrm{tri}}(\theta,P)$\\
\hline
 $45$ & $44.39$ & $0.977$ & $0.967$ & $0.983$ & $0.970$ & $0.976$ \\
 $40$ & $40.46$ & $0.972$ & $0.966$ & $0.983$ & $0.958$ & $0.964$ \\
 $36$ & $35.56$ & $0.968$ & $0.955$ & $0.977$ & $0.907$ & $0.921$ \\
 $30$ & $29.84$ & $0.975$ & $0.965$ & $0.982$ & $0.837$ & $0.851$ \\
 $25$ & $24.33$ & $0.976$ & $0.971$ & $0.986$ & $0.735$ & $0.753$ \\
 $21$ & $20.35$ & $0.978$ & $0.961$ & $0.986$ & $0.645$ & $0.662$ \\
 $17$ & $16.37$ & $0.984$ & $0.971$ & $0.986$ & $0.528$ & $0.548$ \\
 $10$ & $9.33$ & $0.983$ & $0.981$ & $0.990$ & $0.347$ & $0.369$ \\
 $5$ & $4.30$ & $0.989$ & $0.984$ & $0.992$ & $0.164$ & $0.179$ \\
 $0$ & $0.4$ & $0.995$ & $0.994$ & $0.997$ & $0.035$ & $0.013$ \\
\end{tabular}
\end{ruledtabular}
\end{table}

To obtain the experimental value of $I_{185}$ the three-qubit correlation 
functions are sufficient. For the other inequalities $I_{10}$, $I_{96}$, $I_{99}$
we have to evaluate also two-qubit and one-qubit correlations.  
Correlation functions for only two qubits can be calculated summing coincidence 
counts measured for two orthogonal projections on the third irrelevant qubit, 
e.g.:

\begin{equation}
  \langle A_iB_j\rangle = 
  {S_{AB}^+ - S_{AB}^- \over S_{AB}^+ + S_{AB}^-},
  \label{eq:corrFun2}
\end{equation}
where
\begin{eqnarray*}
  S_{AB}^+ &=& f^{+++} + f^{++-} + f^{--+} + f^{---}, \\
  S_{AB}^- &=& f^{+-+} + f^{+--} + f^{-++} + f^{-+-}.
\end{eqnarray*}

And similarly, we can measure the one-qubit correlation from the coincidence 
counts as follows,

\begin{equation}
  \langle A_i\rangle = 
  {S_{A}^+ - S_{A}^- \over S_{A}^+ + S_{A}^-},
  \label{eq:corrFun1}
\end{equation}
where
\begin{eqnarray*}
  S_{A}^+ &=& f^{+++} + f^{++-} + f^{+-+} + f^{+--}, \\
  S_{A}^- &=& f^{-++} + f^{-+-} + f^{--+} + f^{---}.
\end{eqnarray*}

\section{Experimental Results}

\begin{figure}
\centering
\includegraphics[width=8.5cm]{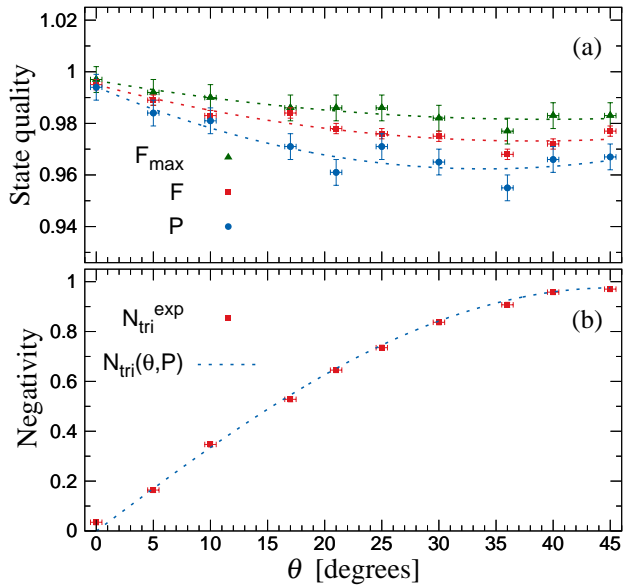}
\caption{(a): Dependence of fidelity $F$, purity $P$, and the upper bound on fidelity $F_{\max}$ as a function of angle $\theta$. Points visualize results calculated from reconstructed states $\rho_{\textrm{exp}}$ by means of quantum state tomography. Dashed lines denote polynomial approximation of $F$, $P$, and $F_{\max}$.
(b): Dependence of tripartite negativity $N^{\textrm{exp}}_{\textrm{tri}} \equiv N_{\textrm{tri}}(\rho_{\textrm{exp}})$ on angle $\theta$. The dashed blue line depicts theoretical prediction $N_{\textrm{tri}}(\theta, P)$, where $P$ corresponds to polynomial approximation of purity presented in (a). The error bars of $N^{\textrm{exp}}_{\textrm{tri}}$ are smaller than $0.005$, i.e., less than the symbol-size.
}
\label{fig:fig_P_N}
\end{figure}

First, we experimentally verify the resulting conclusions presented in Sec. \ref{Sec2} for the prototype GHZ state. 
To do this, we adjust the experimental setup in such a way to have $\theta = (45.0 \pm 0.5)^{\circ}$. This angle is calculated from the experimental measurements as 
\begin{equation}
\theta = \arctan \sqrt{f_{111}/f_{000}}, 
\label{eq:thetaM}
\end{equation} 
where $f_i$ denotes two-photon coincidence rates when projecting all three qubits on logical states $|111\rangle$ and $|000\rangle$, respectively. 

\begin{table}
\caption{Theoretical and experimental values of four Bell-type inequalities. $\tilde{p} I^{\max}_{i} \equiv \tilde{p} \times I^{\max}_{i}(|\mathcal{G}\rangle)$ corresponds to theoretical prediction which includes presence of depolarization $\tilde{p}=\sqrt{\frac{8 P-1}{7}}$ and $I^{\textrm{exp}}_{i} \equiv I^{\max}_{i}(\rho_{\textrm{exp}})$ denotes the maximal attainable values of inequalities $I_{i}$ for the reconstructed state $\rho_{\textrm{exp}}$.
}
\label{tab:tab2}
\begin{ruledtabular}
\begin{tabular}{c c c c c c c c}
 $\theta$  & $\tilde{p} I^{\max}_{10}$ & $I^{\textrm{exp}}_{10}$ & $\tilde{p} I^{\max}_{96}$ & $I^{\textrm{exp}}_{96}$ & $I^{\textrm{exp}}_{99}$ & $\tilde{p} I^{\max}_{185}$ & $I^{\textrm{exp}}_{185}$ \\
\hline
 $45$ & $0.990$ & $0.995$ & $1.252$ & $1.258$ & $1.258$ & $1.387$ & $1.372$\\
 $40$ & $1.040$ & $1.032$ & $1.246$ & $1.252$ & $1.252$ & $1.369$ & $1.355$ \\
 $36$ & $1.064$ & $1.068$ & $1.221$ & $1.230$ & $1.230$ & $1.311$ & $1.283$ \\
 $30$ & $1.089$ & $1.096$ & $1.187$ & $1.199$ & $1.199$ & $1.204$ & $1.185$ \\
 $25$ & $1.102$ & $1.101$ & $1.155$ & $1.158$ & $1.158$ & $1.071$ & $1.040$ \\
 $21$ & $1.087$ & $1.086$ & $1.115$ & $1.117$ & $1.117$ & $0.945$ & $0.913$ \\
 $17$ & $1.073$ & $1.071$ & $1.080$ & $1.080$ & $1.080$ & $0.814$ & $0.823$ \\
 $10$ & $1.049$ & $1.037$ & $1.041$ & $1.031$ & $1.031$ & $0.926$ & $0.925$ \\
  $5$ & $1.014$ & $1.000$ & $1.008$ & $0.999$ & $0.999$ & $0.980$ & $0.981$ \\
  $0$ & $0.996$ & $1.000$ & $0.997$ & $0.999$ & $0.999$ & $0.997$ & $0.999$ \\
\end{tabular}
\end{ruledtabular}
\end{table}

To obtain complete information about generated state $\rho_{\textrm{exp}}$, quantum state tomography and maximum-likelihood estimation have been used to reconstruct the output-state density matrix \cite{Jezek_PRA68, PRL100}. Results are shown in Fig. \ref{fig:tomography_GHZ}.
We then determine fidelity $F$ of $\rho_{\textrm{exp}}$ with respect to the ideal state in Eq. \eqref{eq:ghz}, i.e., we use 
\begin{equation}
F=\max_{\theta} \langle \mathcal{G} |\rho| \mathcal{G} \rangle,
\label{eq:fidelity_opt}
\end{equation}
where maximization is done over $\theta \in \langle 0^{\circ}, 45^{\circ} \rangle$. In the ideal case, the fidelity $F$ shall be equal to $1$. In our case, due to experimental imperfections, the observed output-state fidelity $F(\rho_{exp})=0.977 \pm 0.002$ with the optimal angle $\theta_{opt}$ (which provides maximum $F$) equal to $44.39^{\circ}$, which is in line with previous fast estimation of $\theta$. The error bar of the fidelity is determined by Monte Carlo simulations of Poissonian noise distribution.

To investigate whether the reduced fidelity is caused by improper setting of individual components or by depolarization effects, we calculate the state's purity $P= \textrm{Tr}(\rho^2)$. The undesired but coherent state transformations can arise, e.g., from imperfect polarization compensation by means of the controllers at the input of the beam displacer (photon in mode $1$) while the main reason of the depolarization is noise arising from photon-pair emission of the source, which leads to a reduction in the purity $P$ of the output state. By straightforward calculations we find that $P(\rho_{\textrm{exp}})=0.967 \pm 0.005$. 
This value allows us to determine an upper bound $F_{\max}(\rho_{\textrm{exp}}) = \frac{1}{8} (1 + \sqrt{56 P(\rho_{\textrm{exp}})-7}) = 0.983 \pm 0.003$ on the observable fidelity (discussed in Appendix \ref{appendixC}). As we see, the  fidelities $F(\rho_{\textrm{exp}})$ is very close to $F_{\max}$ and hence, is mainly limited by the presence of noise. 

\begin{figure}
\centering
\includegraphics[width=\columnwidth]{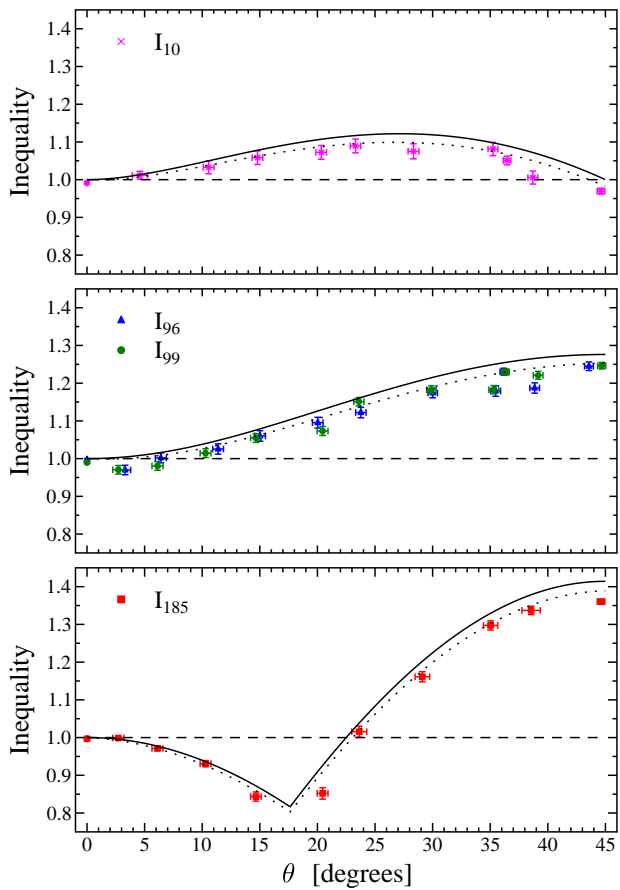}
\caption{Points represent experimental values of normalized inequalities. Solid lines depict theoretical values and dotted lines take into account decrease of the value due to the decreased purity. Dashed line represents the classical limit.
    \label{fig:grafIall}
}
\end{figure}

The influence of parasitic states can be also estimated by means of tripartite negativity $N_{\textrm{tri}} = [N_{A|BC} N_{B|AB} N_{C|AB}]^{1/3}$, where $N_{i|jk}$ indicates the bipartition negativity of qubit $i$ and joined qubits $jk$ \cite{LoveQIP6_2007}. We find that $N_{\textrm{tri}}(\rho_{\textrm{exp}})=0.970 \pm 0.005$, what is in perfect agreement with theoretical predictions, $N_{\textrm{tri}}(45^{\circ}, P)= 0.976 \pm 0.004$, where $P \equiv P(\rho_{\textrm{exp}})$ depicts the purity of $\rho_{\textrm{exp}}$ (see Appendix \ref{appendixC}). 

Next, using the reconstructed state $\rho_{\textrm{exp}}$ and numerical optimization performed with respect to measurements setting, the maximal attainable values of inequalities discussed in Sec. \ref{Sec2} have been determined as
\begin{eqnarray}
I^{\max}_{10}(\rho_{\textrm{exp}}) &=& 0.995 \pm 0.010,\nonumber\\
I^{\max}_{96}(\rho_{\textrm{exp}}) &=& 1.258 \pm 0.009,\nonumber\\
I^{\max}_{99}(\rho_{\textrm{exp}}) &=& 1.258 \pm 0.009,\nonumber\\
I^{\max}_{185}(\rho_{\textrm{exp}}) &=& 1.372 \pm 0.020,
\label{eq:ineq_rec}
\end{eqnarray}
while the results predicted by quantum mechanics for $\theta=45^{\circ}$ are given by $\tilde{p} \times I^{\max}_{10}(|\mathcal{G}\rangle)=0.990 \pm 0.003$, $\tilde{p} \times I^{\max}_{96}(|\mathcal{G}\rangle)=\tilde{p} \times I^{\max}_{99}(|\mathcal{G}\rangle)=1.252 \pm 0.004$, and $\tilde{p} \times I^{\max}_{185}(|\mathcal{G}\rangle)=1.387 \pm 0.004$. Note that theoretical results presented above are corrected by including a small amount of depolarization $\tilde{p}=\sqrt{\frac{8 P(\rho_{\textrm{exp}})-1}{7}}$. As we see, in each case the difference between $I_i(\rho_{\textrm{exp}})$ and theoretical prediction is not greater than $0.015$, which suggests that our measurement outcomes behave as $\tilde{p} \times I^{\max}_{i}(|\mathcal{G}\rangle)$.

Finally, all necessary correlations that enter into $I_{10}$, $I_{96}$, $I_{99}$, and $I_{185}$ are measured directly. To do this we use the correlation function in Eq. \eqref{eq:corrFun} and the appropriate measurements settings described in Sec. \ref{Sec2}. As a result we get
\begin{eqnarray}
I^{\max}_{10} &=& 0.970 \pm 0.008,\nonumber\\
I^{\max}_{96} &=& 1.245 \pm 0.012,\nonumber\\
I^{\max}_{99} &=& 1.246 \pm 0.009,\nonumber\\
I^{\max}_{185} &=& 1.361 \pm 0.006.
\label{eq:ineq_meas}
\end{eqnarray}
Note that the differences between results presented in Eqs. \eqref{eq:ineq_rec} and \eqref{eq:ineq_meas} are not greater than $0.025$. It is also important to emphasize that our result for $I^{\max}_{185}$ is much closer to the maximal attainable value for Svetlichny inequality of $\sqrt{2}$ than the outcomes so far presented in literature, where $I^{\max}_{185}$ is approximately equal to $1.128$ \cite{Lavoienjp11_2009} or $1.115$ \cite{Lupra84_2011a,Lupra84_2011b}.

\begin{figure}
\centering
\includegraphics[width=8.5cm]{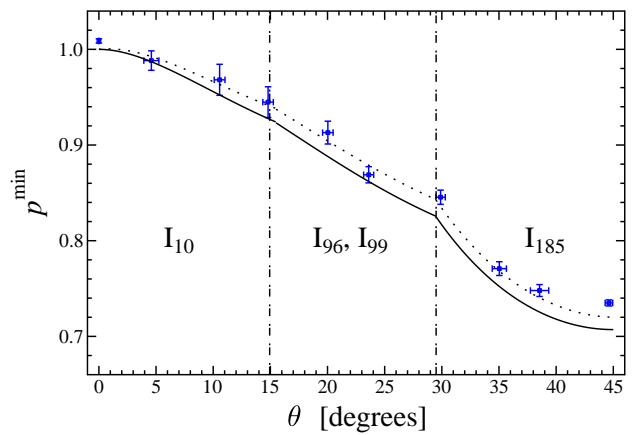}
\caption{Points represent experimental values of $p^{\min}$. Solid line depicts theoretical predictions and dotted line takes into account decrease of the value due to the decreased purity. Dashed-dotted lines split the dependence to three regions, where various Bell-type inequalities yield maximal violation.
\label{fig:grafP}
}
\end{figure}

We repeat our investigation for other angles $\theta$ (presented in Table \ref{tab:tab1}). As we see in Fig. \ref{fig:fig_P_N} (a), for each case the fidelity $F$ (purity $P$) fluctuates between $0.968$ and $0.995$ (between $0.955$ and $0.994$), which confirms a good quality of generated states. Moreover, the tripartite negativity $N_{\textrm{tri}}(\rho)$ is also in perfect agreement with theoretical predictions $N_{\textrm{tri}}(\theta, P)$ (see Fig. \ref{fig:fig_P_N} (b)), implying that the influence of parasitic states is negligible for all analyzed $\theta$. 

Further analysis of the generated states confirms that the maximal attainable value of $I_{10}$, $I_{96}$, $I_{99}$, $I_{185}$ as functions of $\theta$ behaves like $\tilde{p} \times I^{\max}_{10}(|\mathcal{G}\rangle)$ (see Table \ref{tab:tab2}). The maximal difference $\Delta I_i = |I_{i}-\tilde{p} I^{\max}_{i}(|\mathcal{G}\rangle)|$ is given by: $\Delta I_{10} = 0.014$, $\Delta I_{96} = 0.012$, $\Delta I_{99} = 0.012$, $\Delta I_{185} = 0.032$. The same conclusion can be found if one takes direct measurements of Bell-type inequalities into considerations. Our results are presented in Fig. \ref{fig:grafIall}. 
In this case, the highest difference between measurements and theoretical predictions $\tilde{p} \times I^{\max}_{i}(|\mathcal{G}\rangle)$ is given by: $\Delta I_{10} = 0.028$, $\Delta I_{96} = 0.035$, $\Delta I_{99} = 0.030$, $\Delta I_{185} = 0.045$. Note that our measurements of $I_{185}$ are of better quality than the results already published \cite{Lupra84_2011a}, which clearly demonstrate the efficiency of our experimental setup for detection of three-way nonlocality.

Finally, we verify the relation between $p^{\min}$ and angle $\theta$ presented in Eq. \eqref{eq:pminGHZ}. As we see in Fig. \ref{fig:grafP}, due to experimental imperfections all measured values of $p^{\min}$ are greater than theoretical predictions. However, when we include the correction $\tilde{p} \times I^{\max}_{i}(|\mathcal{G}\rangle)$ in Eq. \eqref{eq:pminGHZ}, our results are in perfect agreement with theory (dotted line in Fig. \ref{fig:grafP}).

\section{Conclusions}

In conclusion, we have theoretically and experimentally investigated the violation of four Bell-type inequalities which refer to the  existence of three-way nonlocality. In particular, series of gGHZ states with high fidelity ($F \geq 0.968$) have been prepared experimentally and we have demonstrated the test of the above mentioned Bell-type inequality on these states.
Our results clearly demonstrate that the gGHZ states exhibit nonlocal correlations in the entire range of tripartite entanglement (for any $\theta > 0$), although these states cannot violate the standard Svetlichny inequality, i.e., the three-way nonlocal correlations obtained from these states violate the newly derived Bell-type inequalities based on time-order-dependent local realism models. 
On the basis of that result the minimal value of the $p$ parameter (maximal strength of noise) admissible for the gGHZ states to exhibit genuine multipartite nonlocality has been determined.

Our experiment demonstrates the ability of fast and efficient detection of three-way nonlocality and hence, it opens the door for future investigations of genuine multipartite nonlocal correlations based on the recent concept of nonlocal fraction \cite{Fonsecapra92_2015, Rosierpra96_2017, Lipinskanpj_2018, Barasinskipra98_2018, Fonsecapra98_2018}.

\section{Acknowledgements}
The authors thank Cesnet for providing data management services. A.B. was supported by GA \v{C}R Project No.~17-23005Y, A.Č., J.S. and K.L. by Project No. 17-10003S. Authors also thank M\v{S}MT \v{C}R for support by the project CZ.02.1.01/0.0/0.0/16\_019/0000754. 

\appendix
\section{}
\label[appendix]{appendixA}

In order to derive Eq. \eqref{eq:I96_solution} we adapt the technique used in \cite{Popescupla166_1992}. First, we represent all unit vectors in spherical coordinates, i.e. $\vec{a}_0=(\sin \theta^{a}_{0} \cos \phi^{a}_{0}, \sin \theta^{a}_{0} \sin \phi^{a}_{0}, \cos \theta^{a}_{0})$ and likewise for similarly defined terms. Our aim is to find a set of optimal spherical angles for all measurements $A_0$,...,$C_1$. Due to the  experimental interpretation of measurements $A_0$,...,$C_1$ (see main text) and without lose of generality, we restrict our calculation to azimuthal angles in the interval $[0,\pi]$ and polar angles in $[0, 2 \pi]$. 

Next, let us rewrite Eq. \eqref{eq:ineq96} in terms of unit vectors $\vec{d}_0$ and $\vec{d}_1$ defined such that $\vec{c}_0+\vec{c}_1=2 \vec{d}_0 \cos \eta$ and $\vec{c}_0-\vec{c}_1=2 \vec{d}_1 \sin \eta $. Note that 
\begin{equation}
\vec{d}_0 \cdot \vec{d}_1 = \cos \theta^{d}_{0} \cos \theta^{d}_{1}+\sin \theta^{d}_{0} \sin \theta^{d}_{1} \cos(\phi^{d}_{0}-\phi^{d}_{1})=0.
\label{eq:war1}
\end{equation}
Then, with settings $D_0=\vec{d}_0 \cdot \vec{\sigma}_C$ and $D_1=\vec{d}_1 \cdot \vec{\sigma}_C$ one obtains
\begin{eqnarray}
I_{96} &=& \frac{1}{3} [\langle A_0B_0\rangle - \{\langle A_0D_1\rangle + \langle B_0D_1\rangle\} \sin \eta \nonumber\\
&+& \{\langle D_0\rangle - \langle A_0B_0D_0\rangle + 2 \langle A_1B_1D_0\rangle \} \cos \eta ],\nonumber\\
\label{eq:ineq96d}
\end{eqnarray}
The expectation value of terms in the first parentheses with respect to the gGHZ state is given by
\begin{eqnarray}
\langle A_0D_1\rangle + \langle B_0D_1\rangle\ = -(\cos \theta^{a}_{0} + \cos \theta^{b}_{1}) \cos\theta^{d}_{1}
\end{eqnarray}
and hence, the global maximum of $I_{96}$ with respect to $\theta^{d}_{1}$ is reached when $ \theta^{d}_{1} =\{0,\pi\}$. (It should be noted that for $\eta = \{0,\pi\}$ inequality $I_{96}\leq 1$ for any $\theta^{a}_{0}$, $\theta^{a}_{1}$ etc.) 
Substitution of this result into Eq. \eqref{eq:war1} implies $\theta^{d}_{0} = \pi/2$. Consequently, when inserting $\{\theta^{d}_{0},\theta^{d}_{1}\}=\{\pi/2,\pi\}$ into Eq. \eqref{eq:ineq96d} we find
\begin{eqnarray}
I_{96}(|\mathcal{G}\rangle) &\leq& \frac{1}{3} [\cos \theta^{a}_{0} \cos \theta^{b}_{0}
 + \{\cos \theta^{a}_{0} + \cos \theta^{b}_{0}\} \sin \eta \nonumber\\
&+& \{\cos \phi^{abc}_0 \sin \theta^{a}_{0} \sin \theta^{b}_{0} - 2 \cos \phi^{abc}_1 \sin \theta^{a}_{1} \sin \theta^{b}_{1}\} \nonumber\\
&\times & \cos \eta \sin(2 \theta)],
\label{eq:ineq96d_2}
\end{eqnarray}
where $\phi^{abc}_0 = \phi^{a}_{0} + \phi^{b}_{0} + \phi^{c}_{0}$ and $\phi^{abc}_1 = \phi^{a}_{1} + \phi^{b}_{1} + \phi^{c}_{0}$. Furthermore, due to the symmetry of indexes $a \leftrightarrow b$ one can assume $\theta^{a}_{0}=\theta^{b}_{0}$, $\theta^{a}_{1}=\theta^{b}_{1}$ and likewise for angles $\phi^{a,b}_{i}$. Then, 
\begin{eqnarray}
I_{96}(|\mathcal{G}\rangle) &\leq& \frac{1}{3} [\cos^2 \theta^{a}_{0} 
 + 2 \cos \theta^{a}_{0} \sin \eta \nonumber\\
&+& \{\cos \phi^{abc}_0 \sin^2 \theta^{a}_{0} - 2 \cos \phi^{abc}_1 \sin^2 \theta^{a}_{1}\} \nonumber\\
&\times & \cos \eta \sin(2 \theta)].
\label{eq:ineq96d_3}
\end{eqnarray}
After some standard algebra we find turning points of $I_{96}(|\mathcal{G}\rangle)$ in Eq. \eqref{eq:ineq96d_3} at $\theta^{a}_{0} = 0$, $\theta^{a}_{1} = \pi/2$ and $\phi^{abc}_1 = \pi$ and hence, $I_{96}(|\mathcal{G}\rangle)$ is limited by 
\begin{eqnarray}
I_{96}(|\mathcal{G}\rangle) &\leq& \frac{1}{3} [1 + 2 \sin \eta + 2 \cos \eta \sin(2 \theta)] \nonumber\\
 &\leq & \frac{1}{3} [1 + 2 \sqrt{1+ \sin^2(2 \theta)}],
\label{eq:ineq96d_4}
\end{eqnarray}
where the last inequality comes from $x \cos \alpha + y \sin \alpha \leq \sqrt{x^2 + y^2}$ and is saturated when $\tan \alpha = y/x$. 
Finally, the set of measurement angles that provides the equality in the above expression is
given by $\theta^{a}_{0} = \theta^{b}_{0} = \phi^{c}_{0} = \phi^{c}_{1} = \pi$, $\theta^{a}_{1} = \theta^{b}_{1} = \pi/2$, $\phi^{a}_{0} = \phi^{b}_{0} = \phi^{a}_{1} = \phi^{b}_{1} = 0$ and $\theta^{c}_{0} = \arctan[\sin(2 \theta)]$, $\theta^{c}_{1} = \pi - \arctan[\sin(2 \theta)]$  what ends the proof.

\section{}
\label[appendix]{appendixB}

In order to find a global maximum of the 10th facet inequality a similar method as in Appendix \ref{appendixA} has been applied. In this way the following set of measurements has been determined: 
$\{\theta^{a}_{0},\phi^{a}_{0}\}=\{\pi/2,\pi/3\}$, 
$\{\theta^{a}_{1},\phi^{a}_{1}\}=\{\vartheta_0,-2 \pi/3\}$, 
$\{\theta^{b}_{0},\phi^{b}_{0}\}=\{\vartheta_1,\pi/3\}$, 
$\{\theta^{b}_{1},\phi^{b}_{1}\}=\{\vartheta_1,4 \pi/3\}$,
$\{\theta^{c}_{0},\phi^{b}_{0}\}=\{\vartheta_1,\pi/3\}$, 
$\{\theta^{c}_{1},\phi^{b}_{1}\}=\{\vartheta_1,4 \pi/3\}$, where both $\vartheta_0$ and $\vartheta_1$ depend on angle $\theta$ and satisfy $\partial \tilde{I}_{10} / \partial \vartheta_0=0 \wedge \partial \tilde{I}_{10} / \partial \vartheta_1=0$ with
\begin{eqnarray}
\tilde{I}_{10}&=& \frac{1}{6}[2 \cos \vartheta_1 \{\cos \vartheta_1 + 2 \cos \vartheta_0 \} \nonumber\\
&+& \cos(2 \theta) \{\cos \vartheta_0 - 4 \cos \vartheta_1 + 3 \cos(2 \vartheta_1) \cos \vartheta_0 \} \nonumber\\
&+& 2 \sin^2 \vartheta_1 \sin(2 \theta) \{2 + \sin \vartheta_0\}].
\end{eqnarray}
It should be noted that angles $\vartheta_0$ and $\vartheta_1$ can be easily approximated as: $\vartheta_0 \approx (0.004 \theta^2 - 1.024 \theta + 180)$ and $\vartheta_1 \approx \max\{(\frac{1}{1.6992\cdot10^{-7} + 6.25\cdot10^{-8} \sqrt{\theta}})^{1/3}, (\frac{1}{3.1\cdot10^{-5} + 6.168\cdot10^{-6} \sqrt{\theta}})^{1/2}\}$.

\section{}
\label[appendix]{appendixC}

Let $\rho = p |\psi\rangle \langle \psi| + \frac{1-p}{8} \id$, where $|\psi\rangle$ denotes the tripartite pure state, $\id$ stands for identity matrix, and $0 \leq p \leq 1$. Then, the  fidelity $F = \langle \psi |\rho|\psi\rangle$ becomes
\begin{equation}
F(\rho)= \frac{1+7 p}{8}.
\label{eq:fidelityC1}
\end{equation}
In other words, $F$ increases linearly with $p$. Analogously, the purity is given by 
\begin{equation}
P(\rho)=\textrm{Tr}(\rho^2) = \frac{1+7 p^2}{8}. 
\label{eq:purityC2}
\end{equation}
As we see, when the experimental noise is negligible, both the purity and fidelity should be equal to $1$ and they decrease to $1/8$ if we measure unbiased noise only. 
Moreover, the fidelity decreases with the purity $P$, the presence of undesired states and with any unwanted qubit's rotations around the Bloch sphere, that could be due to, e.g., optical misalignment. 

To get some indication about whether the less-than-unity $F$ is caused mostly due to the purity reduction (assuming the input state $|\psi\rangle$ is pure) or to imperfect preparation of the state (e.g., qubits' rotation), we can estimate the maximal fidelity $F_{\max}$. Specifically, let us assume there are no unwanted rotations, and that the reduction of fidelity comes from the presence of white noise. Then, one can insert the measured value of the purity $P$ in Eq. \eqref{eq:purityC2} into Eq. \eqref{eq:fidelityC1} 
which effectively yields an upper bound to the measured value of the fidelity, 
\begin{equation}
F_{\max}(\rho)=\frac{1}{8} (1+\sqrt{56 P(\rho)-7}).
\end{equation}

Finally, if one assumes that $|\psi\rangle=|\mathcal{G}\rangle$, i.e., $\rho = p |\mathcal{G}\rangle \langle \mathcal{G}|+ \frac{1-p}{8} \id$, then we can easily determine the tripartite negativity $N_{\textrm{tri}}(\rho) = [N_{A|BC}(\rho) N_{B|AB}(\rho) N_{C|AB}(\rho)]^{1/3}$. Here, $N_{i|jk}(\rho) = \frac{||\rho^{\Gamma_i}||-1}{2}$ indicates the bipartition negativity of qubit $i$ and joined qubits $jk$ and $\rho^{\Gamma_i}$ is the partial transpose of $\rho$ with respect to subsystem $i$. After some straightforward calculations one has 
\begin{eqnarray}
N_{\textrm{tri}}(\theta, P) &=& \frac{1}{8} [-1 + \sqrt{\frac{8 P-1}{7}} (1 + 4 \sin(2 \theta)) \nonumber\\
&+& |1 - \sqrt{\frac{8 P-1}{7}} (1 + 4 \sin(2 \theta)) | ],
\end{eqnarray}
where $\theta$ is defined in Eq. \eqref{eq:ghz} and we use the relation $p=\sqrt{\frac{8 P-1}{7}}$ given in Eq. \eqref{eq:purityC2}.


\end{document}